\documentclass[journal]{IEEEtran}

\usepackage{amsmath}
\usepackage{graphicx}
\usepackage{booktabs} 
\usepackage{fullpage}
\usepackage{times}
\usepackage{fancyhdr,graphicx,amsmath,amssymb}
\usepackage[ruled,vlined]{algorithm2e}
\include{pythonlisting}
\usepackage{amsmath, amssymb}
\usepackage{makecell}
\usepackage{multirow}
\usepackage{enumitem}
\usepackage{xcolor}

\begin{document}

\title{Enhanced Microgrid Power Flow Incorporating Hierarchical Control}

\author{Fei~Feng,~\IEEEmembership{Student Member,~IEEE} and
        Peng~Zhang,~\IEEEmembership{Senior Member,~IEEE}
\thanks{This work was supported in part by the National Science Foundation under Grant ECCS-1831811.}
\thanks{The authors are with the Department of Electrical and Computer Engineering, University of Connecticut, Storrs, CT 06269 USA (e-mail: peng.zhang@uconn.edu).}
}

\markboth{}
{Shell \MakeLowercase{\textit{ et al.}}:  Bare Demo of IEEEtran.cls for IEEE Journals}

\maketitle

\begin{abstract}
An enhanced microgrid power flow (EMPF) is devised to incorporate hierarchical control effects. The new contributions are threefold: 1) an advanced-hierarchical-control-based Newton approach is established to accurately assess power sharing and voltage regulation effects; 2) a modified Jacobian matrix is derived to incorporate droop control and various secondary control modes; and 3) the secondary adjustment is calculated on top of the droop-control-based power flow results to ensure a robust Newton solution. Case studies validate that EMPF is efficacious and efficient and can serve as a powerful tool for microgrid operation and monitoring, especially for those highly meshed microgrids in urban areas. 
\end{abstract}

\begin{IEEEkeywords}
Hierarchical control, meshed microgrid, power flow, secondary control.
\end{IEEEkeywords}
\IEEEpeerreviewmaketitle
\section{Introduction}
\IEEEPARstart{M}{icrogrid} has proved to be effective in ensuring electricity resiliency for customers. A most important and indispensable foundation for microgrid operation and management is the power flow analysis~\cite{8309382}. 

However, power flow of islanded microgrid has yet to be addressed because: 
 1) a swing bus no longer exists, rather 2) distributed energy resources (DERs) are operated by hierarchical controls, and 3) microgrid is subject to frequently changes in structure and operating modes~\cite{8636509}. Although modified backward/forward sweep methods~\cite{8309382,8636509,7283640} and Newton method~\cite{7350239} are developed to consider droops in DERs, they fail to handle either meshed microgrids or secondary controllers equipped for frequency and voltage recovery. 

This letter devises an enhanced Newton-type microgrid power flow (EMPF) which fully adapts to both meshed and radial structures.  The main contributions of EMPF lie in : 1) an augmented Newton type formulation of microgrid power flow which supports plug-and-play and allows future extensions into networked microgrids power flow as well as 2) a new Jacobian matrix formulation which is able to incorporate hierarchical control effects and thus precisely considers power sharing and voltage regulation in a modular fashion.
\section{Enhanced Microgrid Power Flow}


In EMPF, in addition to the traditional PV and PQ buses, we introduce a bus type called \emph{DER buses} to which those DERs equipped with droop and/or secondary control are connected. Generally, a slack bus no longer exists because none of the DERs in the droop-based microgrids is able to provide constant voltage and frequency. 
  We can pick an arbitrary DER bus and use its voltage angle as the reference for the rest of the buses.

\vspace{-10pt}

\subsection{EMPF  Formulation}
For an ${N}$-bus microgrid with $\zeta$ DER buses , the power injections from DERs are determined by a two-layer hierarchical control system
~\cite{7112129}. Considering PV, PQ and DER buses, we can derive the EMPF power flow equations as follows 
\begin{equation}\label{E1}
\begin{aligned}
  \mathbf{F}(\boldsymbol{\theta},\mathbf{V},{f})&=\begin{bmatrix}
\mathbf {S}(\mathbf V,f)^\mathbf{G} - \mathbf {S}^\mathbf{L}-\mathbf{\Bar{Y}(\boldsymbol{\theta})} \cdot \mathbf{V} \circ \mathbf{V}  \\[0.3em]
 \mathbf{P}(f)^\mathbf{Gs} - \mathbf {P}^\mathbf{sum}
\end{bmatrix}
   \end{aligned}
  \end{equation}

\noindent where $\mathbf {S}(\mathbf V,f)^\mathbf{G}=[\mathbf{{P}}(f)^\mathbf{G} ,\mathbf{{Q}(V)^\mathbf{G}}]^T\in \mathbb{R}^{(2N-1)\times 1}$ and $\mathbf{{S}^L}=[\mathbf{{P}^L} ,\mathbf{{Q}^L}]^T\in \mathbb{R}^{(2N-1)\times 1}$  are  the generation and load matrices, respectively, $\mathbf{P}(f)^\mathbf{Gs}$ is the total real power from generators, $\circ$ means Hadamard product, $\mathbf {P}^\mathbf{sum}$ is the sum of real power consumption including load and losses. Different from traditional power flow, 
frequency $f$ is a variable in the EMPF formulation.  
 $ \mathbf{\Bar{Y}(\boldsymbol{\theta})}\in \mathbb{R}^{(2N-1)\times N}$ is the extended admittance matrix defined as
\begin{equation}\label{E2}
   \begin{aligned}
  \mathbf{\Bar Y}(\boldsymbol{\theta})&=\begin{bmatrix}
\begin{vmatrix}
 \mathbf{Y}_{ij}
\end{vmatrix}
cos(\theta_i - \theta_j - \alpha_{ij})  \\[0.5em]
\begin{vmatrix}
 \mathbf{Y}_{ij}
\end{vmatrix}
sin(\theta_i - \theta_j - \alpha_{ij})
\end{bmatrix} & i,j\in N
   \end{aligned}
  \end{equation}
\noindent where $\boldsymbol{\theta} \in \mathbb{R}^{(N-1)\times 1}$ is a voltage angle matrix, $\boldsymbol{\alpha_{ij}}$ is the admittance angle of branch $i-j$, 
\vspace{-10pt}
\subsection{Modified Jacobian Matrix}
The modified Jacobian matrix $\mathbf{J}\in \mathbb{R}^{2N\times 2N}$ that incorporates DER behaviors under hierarchical control can be derived from Equation (\ref{E1}), as follows
\vspace{-5pt}
\begin{equation}\label{E3}
   \begin{aligned}
  \mathbf{J}&=\begin{bmatrix}
\frac{\partial \mathbf{F}(\boldsymbol{\theta},\mathbf{V},{f})}{\partial \boldsymbol{\theta}}  ,\frac{\partial \mathbf{F}(\boldsymbol{\theta},\mathbf{V},{f})}{\partial \mathbf{V}}, \frac{\partial \mathbf{F}(\boldsymbol{\theta},\mathbf{V},{f})}{\partial {f}}
\end{bmatrix}
   \end{aligned}
  \end{equation}
where 
\begin{equation}\label{E4}
   \begin{aligned}
  \frac{\partial \mathbf{F}(\boldsymbol{\theta},\mathbf{V},{f})}{\partial \boldsymbol{\theta}}&=\begin{bmatrix}
-\frac{\partial{\mathbf{\Bar Y}(\boldsymbol{\theta})} \cdot \mathbf{V}\circ \mathbf{V}}{\partial \boldsymbol{\theta}}  , \mathbf{0}
\end{bmatrix}^T
   \end{aligned}
  \end{equation}
  \vspace{-10pt}
  \begin{equation}\label{E5}
    \begin{aligned}
  \frac{\partial \mathbf{F}(\boldsymbol{\theta},\mathbf{V},{f})}{\partial \mathbf{V}}&=\begin{bmatrix}\frac{\partial\mathbf {S}(\mathbf V,f)^\mathbf{G}}{\partial \mathbf{V}}-
\frac{{\mathbf{\Bar Y}(\boldsymbol{\theta})} \cdot \partial \mathbf{V}\circ \mathbf{V}}{\partial \mathbf{V}}-\frac{{\mathbf{\Bar Y}(\boldsymbol{\theta})} \cdot \mathbf{V}\circ\partial  \mathbf{V}}{\partial\mathbf{V}}  , \mathbf{0}
\end{bmatrix}^T
   \end{aligned}
  \end{equation}
  \vspace{-10pt}
  \begin{equation}\label{E6}
   \begin{aligned}
  \frac{\partial \mathbf{F}(\boldsymbol{\theta},\mathbf{V},{f})}{\partial {f}}&=\begin{bmatrix}
\frac{\partial\mathbf {S}(\mathbf V,f)^\mathbf{G}} {\partial {f}}  ,\frac{\partial\mathbf{P}(f)^\mathbf{Gs}} {\partial {f}}
\end{bmatrix}^T
   \end{aligned}
  \end{equation}

   Here, the elements in $\mathbf{J}$ matrix are functions of different control modes.  For the droop control mode, the P/F and Q/V droop coefficients are defined as $\mathbf{m}\in \mathbb{R}^{\zeta\times 1} $, $ \mathbf{n}\in \mathbb{R}^{\zeta\times 1} $ respectively. Real power sharing among DERs are achieved through the P/F droop control, as shown in Equation(\ref{E7}-\ref{E8}).

  \begin{equation}\label{E7}
   \begin{aligned}
 \frac{\partial\mathbf {S}(\mathbf V,f)^\mathbf{G}}{\partial {f}}=
\begin{cases}
-\frac{1}{ m_i}, & \mbox{for DER bus} \\
0, & \mbox{otherwise}
\end{cases}
   \end{aligned}
  \end{equation}
     \vspace{-5pt}
  
  \begin{equation}\label{E8}
   \begin{aligned}
 \frac{\partial\mathbf{P}(f)^\mathbf{Gs}} {\partial {f}}=\sum_{i=1}^\zeta -\frac{1}{ m_i}
   \end{aligned}
  \end{equation}
  \vspace{-5pt}
  
   The DER behaviors and corresponding J elements under three typical secondary control modes ~\cite{7112129}  are expressed below: 

\subsubsection  {Reactive Power Sharing Mode (RPS)}  RPS aims to realize proportional reactive power sharing, where the var injection from a leader bus $ Q_1 $ is updated through Q/V droop control and the rest of DER buses follow. Mathematically, the var outputs of DER buses and the corresponding $\mathbf{J}$ elements are 
  \vspace{-2pt}
\begin{equation}\label{E9}
   \begin{aligned}
  \mathbf{Q_{DER}}= \begin{bmatrix}
{Q_{1}( V_{1})}, \rho \cdot \mathbf{Q_F}^*
\end{bmatrix}^T  
   \end{aligned}
  \end{equation}
    \vspace{-10pt}
  {
  \begin{equation}\label{E10}
   \begin{aligned}
 \frac{\partial\mathbf {S}(\mathbf V,f)^\mathbf{G}}{\partial \mathbf{V}}=
\begin{cases}
-\frac{1}{ n_1}, & \mbox{for leader DER bus} \\
0, & \mbox{otherwise}
\end{cases}
   \end{aligned}
  \end{equation}}
   \vspace{-10pt}
   
\noindent where, $\rho$ is the reactive power ratio defined by $ {Q}_{1}/Q_1^*$, and $\mathbf {Q_F}^*$ denotes the rated var outputs of follower buses.
\subsubsection{Voltage Regulation Mode (VR)}
 {VR mode aims to recover the DER bus voltages to their rated values by adjusting the DER reactive power injections.  Thus, }the var outputs of DER buses and the corresponding $\mathbf{J}$ elements are updated by
\begin{equation}\label{E11}
  \begin{aligned}
  \mathbf{Q_{DER}}= diag(\mathbf{V})\cdot diag(\mathbf{Z}_d^{-1})\cdot (\mathbf{V}_d+\mathbf{V}^*-2\mathbf{V}) +\mathbf{Q}_0
   \end{aligned}
  \end{equation}
 \vspace{-10pt}
    {
   \begin{equation}\label{E12}
   \begin{aligned}
 \frac{\partial\mathbf {S}(\mathbf V,f)^\mathbf{G}}{\partial \mathbf{V}}=
\begin{cases}
({Z}_d^{-1})({V}_d+{V}^*-4{V}), & \mbox{for DER bus} \\
0, & \mbox{otherwise}
\end{cases}
   \end{aligned}
  \end{equation}}
    \vspace{-10pt}
    
Similar to~\cite{8309382}, a dummy bus vector with voltages $ \mathbf{V}_d $  is created for DER buses associated with a { sensitivity} vector $ \mathbf{Z}_d $ representing the reactive power differences with respect to the voltage differences between dummy buses and the corresponding DER buses. { Here,} $ \mathbf{V}^* $ denotes rated voltages, and the detailed procedure to update $ \mathbf{V}_d $ can be found in~\cite{8309382}, { $\xi_{\Delta V_d}$ is voltage magnitude error between DER buses and its rated value}. 

\subsubsection{Smart Tuning Mode (ST)}
 { The leader DER bus follows the VR mode to recover back to its rated value, while other DER buses are adjusted for proportional reactive power sharing. Therefore, in this mode, the leader DER bus var output and corresponding $\mathbf{J}$ elements follow Equations (11-12) whereas the rest of DER buses follow Equations (9-10).} 
 

Once $\mathbf{J}$ and $\Delta \mathbf{F}$ are evaluated at the end of each iteration, the microgrid variables $\boldsymbol{\theta}$,$\mathbf{V}$, ${f}$ can be updated for the next iteration by solving the following equation 
  \begin{equation}\label{E13}
   \begin{aligned}
  \Delta \mathbf{F}(\boldsymbol{\theta},\mathbf{V},{f}) &=\mathbf{J} \cdot \begin{bmatrix}
\Delta\boldsymbol{\theta},\Delta\mathbf{V},\Delta{f}
\end{bmatrix}^T
   \end{aligned}
  \end{equation}
The EMPF iterations continue until the errors in those variables reaches the tolerance $\xi$. See Algorithm~\ref{EMPF} for the EMPF pseudo code.

The Newton-type power flow is sensitive to the starting point and relies on {high-quality} initial values for a fast convergence. To ensure the robustness of EMPF incorporating the hierarchical control, it is initialized by the values obtained by running a power flow with droop controls only. Once the convergence criterion is satisfied, all the voltages and branch power flows can be obtained. Because no assumption of microgrid architectures is utilized in EMPF, it can be used to solve power flows for {  arbitrary types of microgrids  such as  radial, meshed, or honeycomb configurations.  }
\vspace{-6pt}
\begin{algorithm}
\SetAlgoLined
  \textbf{{In}itialize:} $\boldsymbol{\theta}$, $\mathbf{V}$, ${f}$, $\xi$, $\rho$(RPS/ST), $\mathbf{V}_d$(VR/ST), $\mathbf{Z}_d$(VR/ST)\;
 \While{$\Delta\boldsymbol{\theta}$, $\Delta\mathbf{V}$,$\Delta{\rho}$,$\Delta{V_d}$, $\Delta{f}$\(\geq\)$\xi $ }
 {
  \eIf{DER bus}{
   Update: $\mathbf{S}(\mathbf V,f)^\mathbf{G}$, $\mathbf{F}(\boldsymbol{\theta},\mathbf{V},{f})$ \textbf{Eq.}(\ref{E1},\ref{E2},\ref{E9},11)\;
   }{
   Update: $\mathbf{F}(\boldsymbol{\theta},\mathbf{V},{f})$ \textbf{Eq.}~(\ref{E1},\ref{E2})\;
  }
  Update: $\mathbf{J}$, $\Delta\boldsymbol{\theta}$, $\Delta\mathbf{V}$, $\Delta{f}$, $\mathbf{P}(f)^\mathbf{Gs}$, $\mathbf {P}^\mathbf{sum}$, $\rho$(RPS/ST), $\mathbf{V}_d$(VR/ST) \textbf{Eq.}(\ref{E3}-\ref{E6},7,\ref{E8},\ref{E10},12)\;
  Update: $\boldsymbol{\theta}$, $\mathbf{V}$, ${f}$\;
 }
 \KwResult{$\boldsymbol{\theta}$, $\mathbf{V}$, ${f}$ and the branch power flow. }
 \caption{EMPF Algorithm}
 \label{EMPF}
\end{algorithm}
\vspace{-10pt}

\section{Case Study}
\begin{figure}[ht]
  \centering
  \includegraphics[width=0.43\textwidth, height=0.28\textwidth]{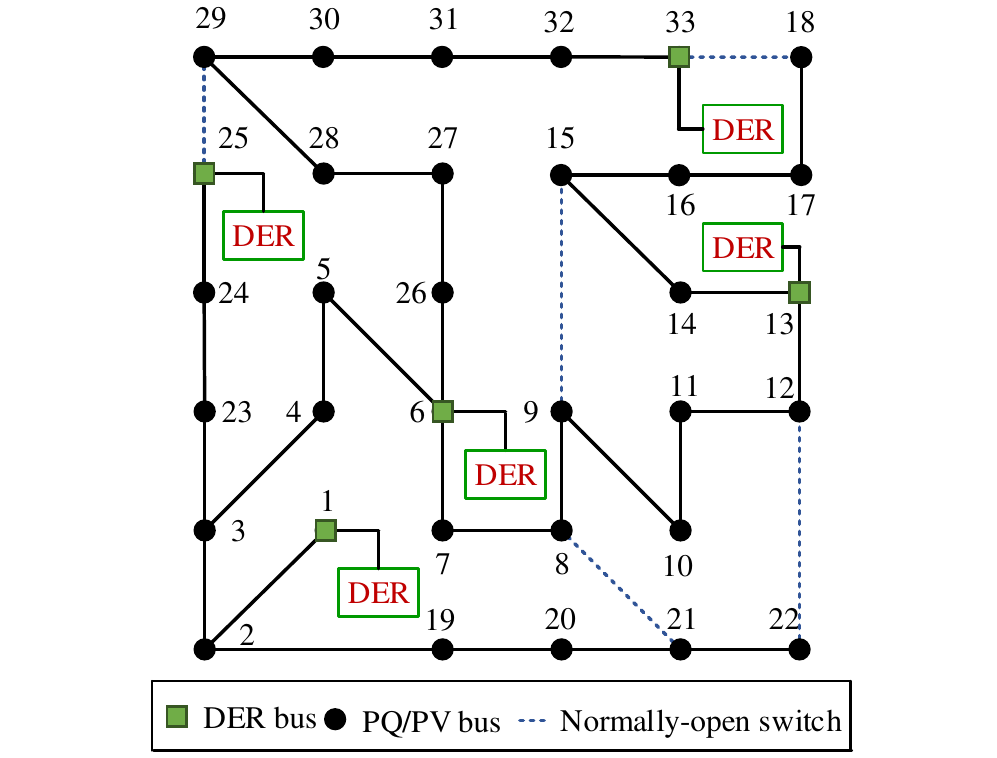}
  \caption{The 33-bus islanded microgrid with 5 DERs}
  \label{F1}
  \vspace{-0.5cm}
\end{figure}

The effectiveness of EMPF is verified on a 33-bus microgrid with 5 DERs (see Fig.~\ref{F1}). For comparison purposes, all system parameters are adopted from~\cite{8309382} except that $\mathbf{Z}_{d} $ = 0.001. By flipping the five normally-open switches, the microgrid configuration can be toggled from radial to meshed one. EMPF calculations are then performed on the radial microgrid (Test I) and the meshed microgrid (Test II). EMPF is implemented in Matlab on a 64-bit, 2.50 GHz PC.  
\vspace{-6pt}
\vspace{-6pt}
\subsection{EMPF Results for Different Microgrid Configurations}
\begin{figure}[ht]
  \centering
  \includegraphics[width=0.39\textwidth]{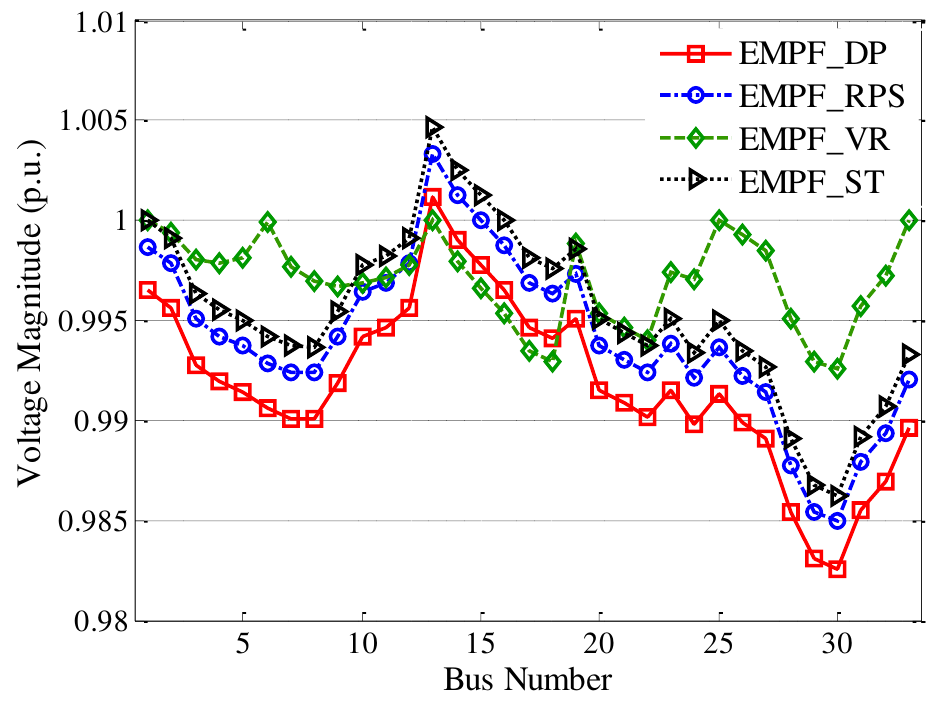}
  \vspace{-0.2cm}
  \caption{Test I: Voltage magnitudes of radial microgrid 
  }
  \label{F2}
\end{figure}

\begin{figure}[ht]
  \centering
  \includegraphics[width=0.39\textwidth]{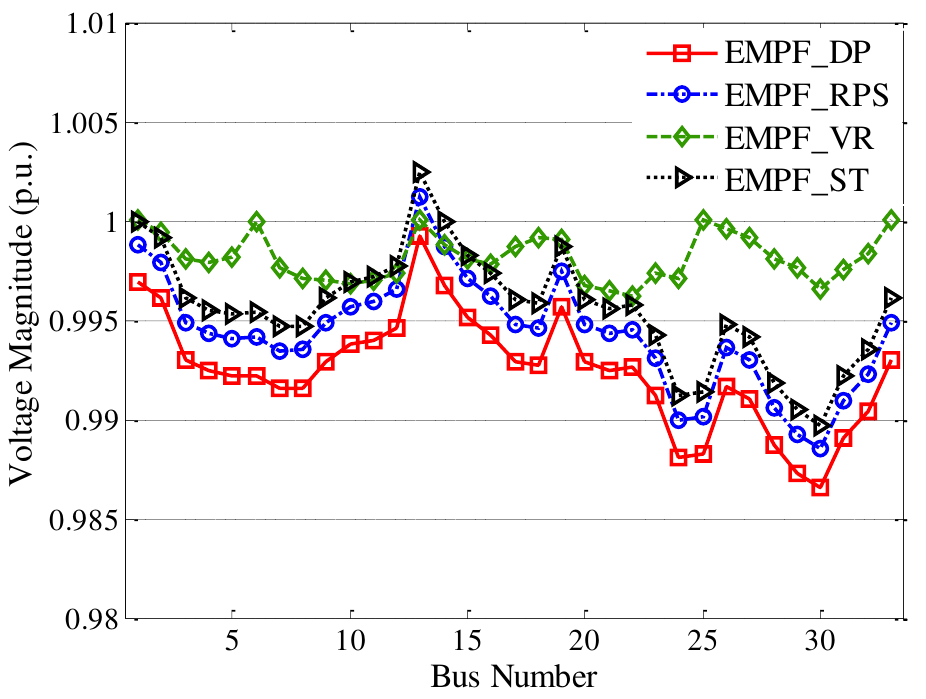}
 \vspace{-0.2cm}
  \caption{Test II: Voltage magnitudes of meshed microgrid 
  }
  \label{F3}
\end{figure}

Voltages obtained from Tests I and II are shown in Figs.~\ref{F2} and~\ref{F3}, respectively. It can be observed that
\begin{itemize}[leftmargin=*]
\item Results in Test I (radial microgrid) are identical to those in~\cite{8309382}, which validates the correctness of EMPF. 
\item Generally, voltages in the meshed microgrid are smoother than those in the radial system. For instance, in the droop mode (EMPF\_DP), the voltage at bus 30 in the meshed system is 0.41\% higher than that in the radial system. This is because DER 25, once the switch 25-29 is closed, will help boost the voltages at neighboring buses including buses 26-33. 
\item Under EMPF\_DP, however, the voltage at DER 13 in the meshed microgrid is lower than that in its radial counterpart because DER 13 has to supply heavy loads at buses 7 and 8 after the switches between 22--12 and 9--15 are closed. 
\end{itemize}

  \vspace{-6pt}
  \vspace{-6pt}
  \vspace{-6pt}
  \vspace{-15pt}

\begin{table}
  \caption{Power Injections from DERs ({\MakeLowercase{p.u.} })} 
  \vspace{-5pt}
  \centering
\begin{tabular}{c c c c c c }
\toprule
   Test & DER\# & DP & RPS & VR & ST\\
  \midrule 
 \multirow{5}*{I} & 1  & 2.50+0.97$i$ & 2.50+0.93$i$ & 2.50-0.90$i$ & 2.50+0.93$i$ \\
   &6  & 0.98+0.91$i$ & 0.98+0.93$i$ & 0.98+2.99$i$ & 0.98+0.93$i$ \\
  & 13 & 1.70+0.89$i$ & 1.70+0.93$i$ & 1.70+0.01$i$ & 1.70+0.93$i$ \\
   &25 & 0.98+0.91$i$ & 0.98+0.93$i$ & 0.98+1.55$i$ & 0.98+0.93$i$ \\
  & 33 & 1.30+0.95$i$ & 1.30+0.93$i$ & 1.30+0.99$i$ & 1.30+0.93$i$ \\
  \midrule
  \multirow{5}*{II} & 1  & 2.50+0.96$i$ & 2.50+0.92$i$ & 2.50-1.18$i$ & 2.50+0.92$i$\\
  & 6  & 0.98+0.91$i$ & 0.98+0.92$i$ & 0.98+2.13$i$ & 0.98+0.92$i$\\
  & 13 & 1.70+0.91$i$ & 1.70+0.92$i$ & 1.70-0.22$i$ & 1.70+0.92$i$\\
  & 25 & 0.98+0.91$i$ & 0.98+0.92$i$ & 0.98+3.08$i$ & 0.98+0.92$i$\\
  & 33 & 1.30+0.94$i$ & 1.30+0.92$i$ & 1.30+0.93$i$ & 1.30+0.92$i$\\
  \bottomrule
\end{tabular}
\label{TableDER}
\vspace{-7pt}
\end{table}
\begin{table}[]
\centering
\caption{CPU time and iteration numbers}
\label{table2}
\vspace{-5pt}
\begin{tabular}{@{}ccccc@{}}
\toprule
Parameter  & DP(I)/(II)& RPS(I)/(II) & VR(I)/(II)  & ST(I)/(II)\\ \midrule
CPU Time(s)& 0.50/0.48 & 0.55/0.54   & 0.82/0.87   & 0.80/0.83 \\
Iteration  & 5/4       & 10/10       & 16/16       & 15/15       \\\bottomrule
\end{tabular}
\end{table}

\subsection{EMPF Results under Various Control Modes}
Table~\ref{TableDER} summarizes DER power injections for both the radial and meshed microgrids under the four control modes. The following insights can be obtained  
\begin{itemize}[leftmargin=*]
\item In generally, microgrid voltage profiles are improved by applying the secondary control, compared with those with droop control only. {For instance, bus 27 voltage under the VR control is 0.9981 which is close to its rated value and is 1.44\% better than that under DP mode only.}
\item {In the RPS mode, the var injections from all DERs are equal because the follower buses share the same reactive power ratio with the leader bus. For instance, in Test I, the var injections of follower DERs 6, 13, 25 and 33 are 0.93 p.u. (base power: 500 kVA) which are equal to the var contribution from the leader bus 1. Therefore, EMPF can realize the proportional reactive power sharing.}
\item { In the ST mode, the leader bus is controlled to fully restore its voltage, as shown in Figures 2 and 3. Meanwhile, the var contribution of each DERs is 0.93 p.u. and 0.92 p.u. for the radial and meshed microgrids, respectively, because in this mode the follower buses still follow the RPS mode.}
\item {In the VR mode, the voltages at DER buses can be recovered to the nominal values. However, compared with the RPS and ST modes, it often leads to irregular power sharing among DERs. Therefore, it indicates that the VR mode is only feasible when DERs have adequate reactive power capacity.}
\end{itemize}

{Please note that EMPF is different from the microgrid power flow approach in paper [1] which is based on the modified backward/forward sweep and thus limited to dealing with a radially structured microgrid. Our method, instead, is based on an augmented, plug-and-play Newton approach that can handle all possible microgrid configurations effectively. Even for the radial system analysis, our method has also shown some better performance. For instance, in VR mode, EMPF iterates only 16 times ($\xi_{\Delta f,~\Delta \rho}=10 ^{-3}$, $\xi_{\Delta V_d}=10 ^{-4}$, $\xi_{\Delta V,~\Delta \theta}=10 ^{-5}$), whereas it takes the method in [1] 173 iterations ($\epsilon_1=10 ^{-3}$, $\epsilon_2=10 ^{-3}$, $\epsilon_3=10 ^{-4}$) to converge. Another desirable feature of EMPF is that there is no limit in selecting the sensitivity $\mathbf{Z}_{d} $ as the value of $\mathbf{Z}_{d} $ does not affect the convergence performance. }
  
\section{Conclusion}
EMPF is developed to accurately calculate power flow in microgrids equipped with hierarchical control. {  Test results exhibit that EMPF can be used for both radial and meshed microgrids. Excellent convergence performance of EMPF demonstrates its efficacy and scalability.} EMPF can be implemented as an essential functionality in microgrid energy management systems and can also be used to provide accurate initial values for microgrid stability and security studies. Next, it will be generalized for { power flow calculations} in networked microgrids.

\bibliographystyle{IEEEtran}
\bibliography{ref}

\end{document}